\documentclass[preprint,12pt]{elsarticle}

\usepackage{graphicx}
\usepackage{amssymb}
\usepackage{lineno}
\usepackage{color}
\usepackage{wrapfig}
\usepackage{amsmath}
\usepackage{xcolor}
\usepackage{adjustbox}
\usepackage{subcaption}
\usepackage{tabularx}
\usepackage{float}
\usepackage{hyperref}
\usepackage{url}
\usepackage{doi}
\usepackage{caption}

\usepackage[titletoc, page]{appendix}
\biboptions{sort&compress}
\biboptions{semicolon}

\usepackage{natbib}
\setcitestyle{authoryear,open={(},close={)}}

\journal{Biostatistics}

\begin{document}

\begin{frontmatter}

\title{Geographically-dependent individual-level models for infectious diseases transmission}

\author[1]{Md Mahsin}
\author[1,2]{Rob Deardon}
\author[3]{Patrick Brown}

\address[1]{ Department of Mathematics and Statistics, University of Calgary, Canada }
\address[2]{Department of Production Animal Health, Faculty of Veterinary Medicine, University of Calgary, Canada }
\address[3]{ Department of Statistical Sciences, University of Toronto, Canada }

\begin{abstract}
Infectious disease models can be of great use for understanding the underlying mechanisms that influence the spread of diseases and predicting future disease progression. Modeling has been increasingly used to evaluate the potential impact of different control measures and to guide public health policy decisions.  In recent years, there has been rapid progress in developing spatio-temporal modeling of infectious diseases and an example of such recent developments is the discrete time individual-level models (ILMs). These models are well developed and provide a common framework for modeling many disease systems, however, they assume the probability of disease transmission between two individuals depends only on their spatial separation and not on their spatial locations. In cases where spatial location itself is important for understanding the spread of emerging infectious diseases and identifying their causes, it would be beneficial to incorporate the effect of spatial location in the model. In this study, we thus generalize the ILMs to a new class of geographically-dependent ILMs (GD-ILMs), to allow for the evaluation of the effect of spatially varying risk factors (e.g., education, social deprivation, environmental), as well as unobserved spatial structure, upon the transmission of infectious disease.  Specifically, we consider a conditional autoregressive (CAR) model to capture the effects of unobserved spatially structured latent covariates or measurement error. This results in flexible infectious disease models that can be used for formulating etiological hypotheses and identifying geographical regions of unusually high risk to formulate preventive action. The reliability of these models are investigated on a combination of simulated epidemic data and Alberta seasonal influenza outbreak data ($2009$). This new class of models is fitted to data within a Bayesian statistical framework using Markov chain Monte Carlo (MCMC) methods.
\end{abstract}


\end{frontmatter}

\section{Introduction}

Infectious disease models can be of great value for understanding the underlying mechanisms that influence the spread of diseases and predicting future disease progression. Modeling has  increasingly been used to evaluate the potential impact of different control measures and to guide public health policy decisions \citep{keeling2011modeling, basu2013complexity}. In particular, model-based surveillance techniques for infectious disease provide better insight into the etiology, spread, prediction, and control of infectious diseases, and so they are widely used for outbreak detection \citep{lawson2016handbook}. In recent years, there has been rapid progress in developing spatio-temporal models of infectious diseases, with particular interest in the development of statistical methods to fit such models to data \citep{ riley2007large, o2010introduction}. Much of this growth can be attributed to the fact that data have become  more refined spatially \citep{gog2014spatial}, and that increased computational computer power allows for more complex models to be investigated.

One example of such recent developments are the discrete time individual-level models (ILMs) of  \citet{deardon2010inference} that can be used to make inference about the spread of a disease through a heterogeneous population. The key feature of an ILM is that it can take into account covariate information on susceptible and infectious individuals (e.g., age, genetics, lifestyle factors), as well as shared covariate information such as  separation distance or contact measures (e.g., sexual partnerships for a human STI, or shared household or workplace). Such ILMs may be used to identify possible risk factors, predict the course of an on-going outbreak, or evaluate control measures.



Our focus here is an ILMs in which spatial mechanisms are of potential importance. Although such spatial ILMs are well developed and provide a common framework for modeling many disease systems, they currently assume the probability of disease transmission between two individuals depends only on their spatial separation and not on their spatial locations.  In cases where spatial location itself is important for understanding the spread of emerging infectious diseases and identifying their causes, it would be beneficial to incorporate the effect of spatial location in the model. In addition, spatially varying demographic and environmental factors could influence the disease transmission. For instance, the transmission of  vector-borne infectious diseases  are highly influenced by spatially correlated environmental factors such as temperature, air quality, rainfall, and humidity \citep{morens2004challenge, palaniyandi2017environmental}. Furthermore, the effects of weather on the transmission rate of pandemic H1N1 influenza in 2009 were investigated in Canada using aggregated cases at each province, and \citet{he2013patterns} found that simulations from models incorporating weather factors were much more in line with observed data than those from models without weather factors.


In this paper, we thus generalize the ILMs of \citet{deardon2010inference} to a new class of geographically-dependent ILMs (GD-ILMs), to allow for the evaluation of the effect of spatially varying risk factors (e.g., social, environmental, topological), as well as unobserved spatial structure, upon the transmission of infectious disease. Specifically, we consider a conditional autoregressive (CAR) model to capture the effects of unobserved spatially structured latent covariates or measurement error. This results in flexible infectious disease models  that can be used for formulating etiological hypotheses and identifying geographical regions of unusually high risk to formulate preventive action. Therefore, this new class of GD-ILMs will provide for better understanding of the spatiotemporal dynamics of disease spread facilitating a greater understanding of the impact of  policies and interventions for controlling epidemic outbreaks. The reliability of these models are investigated on a combination of simulated epidemic data and real data. Specifically, a simulation study is conducted to investigate the performance of our proposed GD-ILMs in terms of their ability to ascertain infectious disease dynamics, both globally and within specific geographical regions of interest. Additionally, we apply the GD-ILMs to real data on the Alberta seasonal influenza outbreak that occurred in $2009$ in Calgary, Canada. This data is modelled at the level of the smallest standardized spatial unit available in Canada, the ``dissemination area" (DA).  Regional effects are modelled at a coarser level, over the sixteen ``local geographic areas" (LGAs) in Calgary.  Of key importance is ascertaining if there is evidence of LGA-level spatial effects, and if spatial effects can be detected regarding transmission between DAs. 

As is typical in infectious disease modeling, this new class of models is fitted to data within a Bayesian statistical framework. More specifically, Markov chain Monte Carlo (MCMC) techniques, utilizing Gibbs sampling and the Metropolis-Hastings algorithm, are used to iteratively sample model parameter estimates from the posterior distribution \citep{gelman2013bayesian}. However, the computational time required to compute the likelihood for the ILMs of \citet{deardon2010inference} is intensive, and increases significantly for epidemics of even moderately large size. Unfortunately, the integrated nested Laplace approximation (INLA) approach, which is commonly used in geostatistical or disease mapping models, cannot be applied here since ILMs and GD-ILMs fall outside the classes of models to which it can be applied \citep{rue2009approximate}. Several approaches have been considered to overcome this computational burden such as kernel linearization \citep{deardon2010inference, kwong2012linearized}, data sampling-based likelihood approximation \citep{malik2016parameterizing}, and Gaussian process emulation \citep{pokharel2016gaussian}. The added complexity inherent in our new class of ILMs has the potential to make this computational bottleneck even worse. To help alleviate this problem we consider the use of a ``region-restricted" GD-ILMs in which disease transmission can only occur in localized areas.

The structure of this paper is as follows. In Section \ref{method}, the general ILMs of \citet{deardon2010inference} are introduced and the new class of GD-ILMs with computational details, are presented. In Section \ref{simulation}, a simulation study is conducted to demonstrate the application of GD-ILMs.  Section \ref{rdata} describes an application of GD-ILMs to Alberta seasonal influenza outbreaks that occurred in $2009$ in Calgary, Canada. We conclude the paper with a discussion in Section \ref{discuss}.

\section{Models and Inference} \label{method}

\citet{deardon2010inference} introduced a class of flexible discrete time stochastic epidemic models known as individual-level models (ILMs), to model the spread of infectious diseases in heterogeneous populations. These models are capable of modeling infectious disease epidemics through space and time at the level of individual units between which the disease is assumed to be spreading in a population. These units could be, for instance, individual persons, animals or plants, or could represent individual houses, farms, schools or regions. The ILM terminology derives from the fact that individual-level covariate information, such as the spatial location of a house, or the vaccination status of individual people, can be incorporated into the ILM framework.

Here, we consider GD-ILMs that fit into a discrete-time susceptible-infectious-removed ($ \mathcal{SIR} $) compartmental framework (e.g., \citealp{anderson1991infectious}). This compartmental structure means that at any given point in time, individuals in the population could be in one of three states: susceptible ($ \mathcal{S}$), infectious ($\mathcal{I}$), or removed ($\mathcal {R}$).  In the first state, individuals do not have the disease but are susceptible  and can contract it, while in the second state, individuals are considered to have contracted the disease and are capable of spreading the infection. In the final state, individuals are assumed to be removed from the population, possibly through recovery with acquired immunity, quarantine or death.  Over the course of the epidemic, individuals either remain susceptible or move through the three states in the order $\mathcal{S} \rightarrow \mathcal{I} \rightarrow \mathcal{R}$. However, GD-ILMs can easily be fitted within other compartmental frameworks (e.g., $\mathcal{SI}$, $\mathcal{SEIR}$). 

\subsection{The General ILM of \citet{deardon2010inference}}

In this section we present the general form of the epidemic ILM based on \citet{deardon2010inference} where the heterogeneity of infectious disease transmissions are allowed at the individual-level. We denote the set of individuals who are susceptible, infectious or removed at time $t$ as $ \mathcal{S}(t)$,  $ \mathcal{I}(t)$ or  $ \mathcal{R}(t)$ respectively. Note, for given $t$, $ \mathcal{S}(t)$,  $ \mathcal{I}(t)$ and  $ \mathcal{R}(t)$ are mutually exclusive.  Here, we assume time is discretized so that time point $t$, for $t = 1, 2, ..., \infty$, represents a continuous time interval $[t, t + 1)$. 

Let $P (i, t)$ be the probability of a susceptible individual $i$ being infected at time $t$. Then a general form of the ILM is given by \citet{deardon2010inference} as:

\begin{equation}\label{ilm}
P (i, t) = 1 - \exp\left [\left\{-\Omega_S(i)\sum_{j \in \mathcal{I}(t)}^{} \Omega_T(j) \kappa(i,j)\right\} - \varepsilon(i, t)\right], 
\end{equation}
where $\mathcal{I}(t)$ is the set of infectious individuals at time $t$. The functions $\Omega_S(i)$ and $\Omega_T(j)$ represent risk factors associated with susceptible individual $i$ contracting, and infectious individual $j$ passing on, the disease, respectively. Risk factors that involve both the infected and susceptible individuals, such as spatial separation or contact networks, are incorporated through the infection kernel, $\kappa(i,j)$. Finally, the sparks term, $\varepsilon(i, t)$, represents infections that are not well explained by the $\Omega_S(i)$, $\Omega_T(j)$, and $\kappa(i,j)$ terms (e.g. infections originating from outside the study population). For example, $\varepsilon(i, t) = \varepsilon$ could be used to represent purely random infections that occur with equal probability throughout the susceptible population at any given time.



\subsection{Geographically-dependent ILMs (GD-ILMs)} \label{gd-ilm}

The ILMs of \citet{deardon2010inference} are generalized to a new class  termed as geographically-dependent individual-level models (GD-ILMs), to allow for evaluating the effect of spatially varying risk factors upon the transmission of infectious disease. These could be social factors (e.g., education, social deprivation), environmental factors (e.g., temperature, air quality, rainfall, and humidity), and / or represent unobserved spatial structure. The GD-ILMs have the form: 

\begin{equation}\label{gdilm}
P (i, k, t) = 1 - \exp\left [\left\{-\Omega_S(i, k)\sum_{j \in \mathcal{I}(t, k, \xi (k))}^{} \Omega_T(j, k) \kappa(i,j)\right\} - \varepsilon(i, k, t)\right],
\end{equation}
where $k$ represents the area index which varies from $1$ to $K$,  $\xi (k)$ is the set of neighboring areas that are adjacent to area $k$, and $\mathcal{I}(t, k, \xi (k))$ is the set of infectious individuals at time $t$ in the $k^{th}$ area and its neighboring areas. We consider areas to be neighbors if they share a common geographical boundary. Here, $\Omega_S(i, k)$ is a susceptibility function of potential risk factors associated with susceptible individual $i$ in area $k$ contracting the disease; $\Omega_T(j, k)$ is a transmissibility function of potential risk factors associated with infectious individual $j$ in area $k$ passing on the disease; $\kappa(i,j)$ is the infection kernel that represents risk factors associated with both susceptible $(i)$ and infectious $(j)$ individuals; and the sparks term, $\varepsilon(i, k, t)$, represents ``random" infections that are not otherwise explained by the model.

\subsubsection{GD-ILMs with Covariates}

The aforementioned susceptibility and transmissibility functions, $\Omega_S(i, k)$ and $\Omega_T(j, k)$, respectively, can be used to model individual-level covariates and area-level covariates of interest. For instance, if the individuals being modeled are humans, we may wish to identify the vulnerable age groups and gender along with the estimation of vaccination effect in the susceptibility function. In addition, the effect of area-level covariates such as temperature, relative humidity, particulate matter, social deprivation index and so on might be interested on the susceptibility function.  In order to account for the effects of covariates on the susceptibility function, let $\mathbf{X} (i, k)$  be the vector of covariates associated with susceptible individual $i$ in area $k$ and $\mathbf{X} (k)$ be the area-level covariates vector corresponding to area $k$. Then, we define the susceptibility function as $\Omega_S(i, k) = \exp\Big(\alpha + \mathbf{X} (i, k)^{\prime}\boldsymbol{\alpha_{1}} +  \mathbf{X} (k)^{\prime}\boldsymbol{\alpha_{2}} +  \mathbf{X} (k, t - \rho )^{\prime}\boldsymbol{\alpha_{3}} + \phi_k\Big)$, where $\alpha > 0$ is a constant infectivity parameter; $\boldsymbol{\alpha_{1}}$, $\boldsymbol{\alpha_{2}}$, and $\boldsymbol{\alpha_{3}}$ are the vector of parameters for individual-level covariates, spatially varying area-level covariates, and area-level covariates associated with environmental factors at time $t$, effect on the susceptible populations while the environmental factors are considered at a lag of $\rho$ time point (e.g. week, day), respectively.  The spatial random effect, $\phi_k$, corresponds to spatially structured heterogeneity and represents spatial variation in the mechanism of the disease spread between areas that captures the effects of unobserved variables with an underlying spatial pattern. In the Bayesian framework, this spatial structure is encoded into the prior distribution for these spatial random effects and involves the definition of relationships between spatially close areas. We consider a {\it conditional autoregressive } (CAR) model to capture the effects of unobserved spatially structured latent covariates or measurement error by the spatial random effect $\phi_k$ (see Section \ref{car}). This approach models the effect of proximity using a first-order neighborhood structure.

We assume that $\Omega_T(j, k) = 1 $ in the the GD-ILMs in (\ref{gdilm}),  that is, the individual-level covariates are not considered in the transmissibility function.  Finally, among a number of well known spatial kernel transmission functions, the power law spatial kernel function is used in this study. It is defined as $\kappa(i,j) = d_{ij}^{- \delta}$, where $\delta > 0$ is a spatial infectivity parameter and $d_{ij}$ is a measure of geographic distance between susceptible individual $i$ and infectious individual $j$. This distance kernel allow the infection rates to decrease when the distance between susceptible and infectious individuals increases. For directly transmitted human diseases, this geographic distance could be the Euclidean distance between the homes or individuals $i$ and $j$.  Euclidean distance-based kernel transmission functions have been greatly used by many authors \citep{ boender2010transmission, parry2014bayesian}. For instance, \citep{savill2006topographic} reported that  Euclidean distance is better predictor of transmission risk than shortest and quickest routes via road, and appropriate to most areas except where major geographical features intervene.

We then define the rate of infectivity to a susceptible individual $i$ at time point $t$ in a given area $k$ based on (\ref{gdilm}) as,
\begin{equation}\label{infrate}
\eta_{i, k}(t) = \exp\Big(\alpha + \mathbf{X} (i, k)^{\prime}\boldsymbol{\alpha_{1}} +  \mathbf{X} (k)^{\prime}\boldsymbol{\alpha_{2}} +  \mathbf{X} (k, t - \rho )^{\prime}\boldsymbol{\alpha_{3}} + \phi_k\Big)\sum_{j \in \mathcal{I}(t, k, \xi (k))}^{}  d_{ij}^{- \delta}. 
\end{equation}

\noindent Therefore, the GD-ILMs given in (\ref{gdilm}) can be written as 
\begin{equation}\label{gdilm2}
 P(i, k, t) = 1 - \exp \Big( - \big(\eta_{i, k}(t) + \varepsilon(i, k, t)\big)\Big). 
\end{equation}
\noindent In these GD-ILMs, we consider the transmission of disease from an infectious individual to a susceptible individual arising into three types of sources: the effects of unobserved spatially structured latent covariates or measurement error, an infectious individual within the same area, and an infectious individual in an adjacent area. For instance, the diseases may be spread out from infected individuals in neighboring areas and a potential source of disease transmission to susceptible individuals for a specific area. In order to better understand the underlying mechanisms of the disease transmission, it might be necessary to incorporate the neighboring areas source of transmission. 

\subsubsection{Conditional autoregressive models}\label{car}

A popular class of models used to represent the spatial random effects, $\phi_k$, is the conditional autoregressive (CAR) model \citep{besag1974spatial}, which are a type of Markov random field model. The spatial dependence is expressed conditionally by requiring that the random effect in a given area, given the values in all other areas, depends only on a small set of neighboring values. The models were extended to a fully Bayesian setting by  \citet{besag1991bayesian} and are readily implemented via  MCMC algorithms. Specification of the CAR models is directly linked to its covariance matrix. This matrix is tremendously important as it allows the incorporation of spatial structure into the CAR model. Different specifications for this matix results in a number of CAR models. In this study, the spatial random effects, $\Phi = (\phi_1, \phi_2, ....,\phi_K)$, are modelled using the so-called LCAR process \citep{leroux2000estimation}. Details of this LCAR model can be found at Section \ref{lcar} of the Appendix.

\subsection{Bayesian Inference}\label{mcmc}

Our parameter estimation is carried out under a Bayesian statistical framework using Markov chain Monte Carlo (MCMC) methods. Assuming known infection and removal times, the likelihood function for the GD-ILMs is the product of all infection and non-infection events over the entire observed epidemic period $\{1, 2, ..., T\}$ and across all areas $\{1, 2, ..., K\}$, and is given by

\begin{equation}\label{likgdilm}
L\big(\Theta; \boldsymbol{\mathcal{S, I, R}}\big) = f(\boldsymbol{\mathcal{S, I, R}}| \Theta) = \prod_{k = 1}^{K}\prod_{t = 1}^{T} f_{t,k} \big(\mathcal{S}(t, k), \mathcal{I}(t, k), \mathcal{R}(t, k)|\Theta \big),
\end{equation}
\noindent where $\boldsymbol{\mathcal{S}} = \left\{\mathcal{S}(t, k)_{k = 1}^{K}\right\}_{t = 1}^{T}$, $\boldsymbol{\mathcal{I}} = \left\{\mathcal{I}(t, k)_{k = 1}^{K}\right\}_{t = 1}^{T}$, $\boldsymbol{\mathcal{R}} = \left\{\mathcal{R}(t, k)_{k = 1}^{K}\right\}_{t = 1}^{T}$ and the joint probability of all new infections occurring in time interval $[t; t + 1)$ for the GD-ILMs is

\begin{equation} \label{gdlik}
f_{t, k} \big( \mathcal{S}(t, k), \mathcal{I}(t, k), \mathcal{R}(t, k)|\Theta \big) = \left[ \prod_{i \in \mathcal{I}(t + 1 ,  k, \xi (k)) \backslash \mathcal{I}(t, k, \xi (k) )}{} P (i, k, t) \right]   \left[ \prod_{i \in \mathcal{S}(t + 1, k )}{} \big(1 - P (i, k, t)\big) \right].
\end{equation}

In combination with a prior density, $\pi(\Theta)$, on our parameter set $\Theta = (\alpha, \boldsymbol{\alpha_{1}}, \boldsymbol{\alpha_{2}}, \boldsymbol{\alpha_{3}}, \delta, \lambda, \sigma^2 )$, we can obtain the posterior distribution, $\pi(\Theta | \boldsymbol{\mathcal{S, I, R}})$, up to a constant of proportionality. To explore the posterior distribution for each of the model parameters, we use a combination of Gibbs sampling \citep{gelman2013bayesian} and  random-walk Metropolis Hastings (RWMH) algorithm \citep{metropolis1953equation, hastings1970monte, chib1995understanding}.

\section{Simulation Study}\label{simulation}
The purpose of this simulation study is to investigate the performance of the proposed GD-ILMs in terms of their ability to ascertain infectious disease dynamics, both globally and within specific regions of interest. For simplicity, here we assume that  $\varepsilon(i, k, t) = 0$ and the infectious period $\gamma$ is assumed to be constant for all individuals.

\subsection{Models for simulation study}
To illustrate the GD-ILMs described above,  we consider epidemic simulations based on the city of Calgary. The city of Calgary consists of $K = 16$ health areas, known as local geographic areas (LGAs), with a total population of $1,239,220 $ in a geographical region of $825.3$ km$^2$ \citep{AHS}. To simulate realizations of an epidemic, we consider two different forms of model: (i) a model that allows disease transmission to occur only within each health area and their neighboring areas (region restricted), $\xi (k)$; and (ii) a model that allows disease transmission to occur between individuals across the whole study area (global). These two models are specified as 
\begin{equation}\label{egdilmsim}
P(i, k, t) = 1 - \exp\left [ - \exp \Big(\alpha + \alpha_{1} X_{i, k} + \phi_k \Big) \sum_{j \in \mathcal{I}(t, k, \xi (k) )}^{} d_{ij}^{- \delta} \right], 
\end{equation}
and 
\begin{equation}\label{gdilm2sim}
P(i, k, t) = 1 - \exp\left [- \exp(\alpha + \alpha_{1} X_{i, k} + \phi_k) \sum_{j \in \mathcal{I}(t)}^{} d_{ij}^{- \delta}\right ],
\end{equation}
respectively, where $X_{i, k}$ is a covariate associated with susceptible individual $i$ in the $k^{th}$ LGA and $\mathcal{I}(t)$ is the set of infectious individuals at time $t$ in the whole study area.

\subsection{Epidemic Simulation}\label{episimu}

Dissemination areas (DAs) are the smallest standard geographic unit in Canada and have an average population of 400 to 700 people \citep{DA}. It is at this level that we consider the spread of an epidemic (i.e. we treat the DAs as our individual units). A total of $1570$ DAs, ranging from  $37$ to $225$, are considered across the $16$ LGAs. The centroid location of each DA and the covariate, the population sizes of each DA from the $2011$ Canadian census, are then used to simulate epidemics using the region restricted (\ref{egdilmsim}), and global (\ref{gdilm2sim}), models under $\mathcal{SIR}$ settings. Three different scenarios for each simulated epidemics are considered, using a fixed set of parameters and varying the number of infected individual DAs at the beginning of the simulation. The values used for the constant infectivity rate, DA population size, and spatial parameters are, $\alpha = 0.30$, $\alpha_1 = 0.40$ and $\delta = 4.0$,  respectively. These parameters were chosen to obtain ``informative" epidemics (i.e. epidemics that tend to proliferate rather than `die out', but also do not spread too quickly). 

\begin{figure}[!h]
\centering
\includegraphics[height = 4.5in, width = 6.5in]{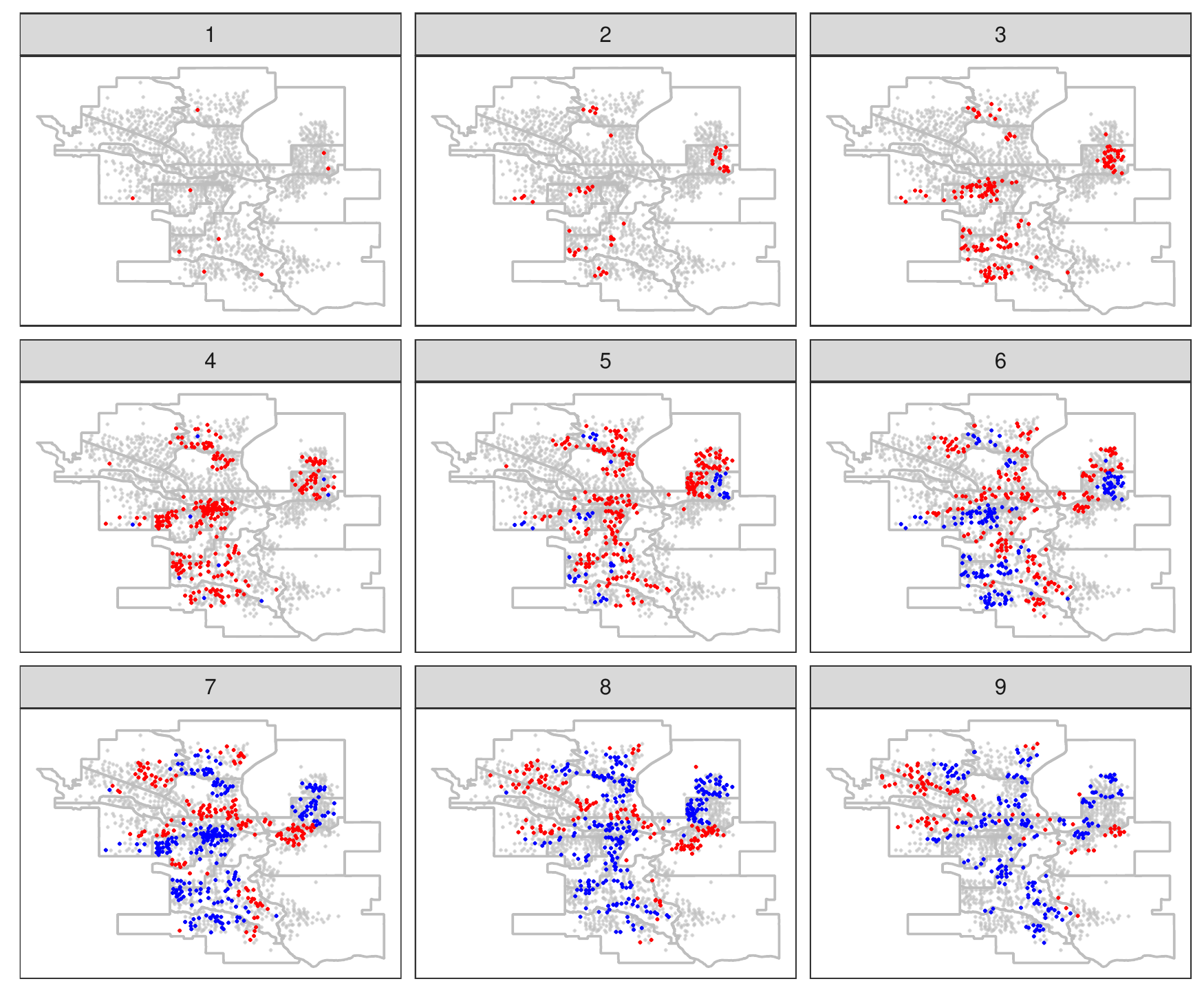}
\caption{A subset of a realization of the epidemic progress maps made from $t = 1$ up to $ t = 9$  across the $16$ LGAs for the city of Calgary, Canada. Newly infected, susceptible, and removal individual DAs are denoted by red, grey and blue dots, respectively.}
\label{figh1}
\end{figure}

The random effects $\Phi = (\phi_1, \phi_2, ..., \phi_{16})$ are generated from a multivariate normal distribution with mean vector $\mathbf{0}$ and  variance hyperparameter, $\sigma = 0.60$ in the covariance matrix. In addition,  we considered three different settings for the spatial dependence hyperparameter $\lambda$: (i) weak spatial dependence ($\lambda = 0.30$); (ii) moderate spatial dependence ($\lambda = 0.50$); and (iii) strong spatial dependence ($\lambda = 0.80$). In each epidemic simulation, a fixed infectious period $\gamma = 3$ is assumed for each DA, and the infectious state of each individual (e.g. $\mathcal{S}, \mathcal{I}$ or $\mathcal{R}$) is recorded at $t = 1, 2, ..., 20$ time points. For each scenario and settings of $\lambda$, we generate a total of $10$ simulated epidemics. The scenarios considered for the simulation study are:

 Scenario $1$ (S$1$): We assume that $9$ individual DAs are infectious at time $t = 1$. For convenience, we choose these nine are the first nine infections observed on the $25^{th}$ October, $2009$ in Calgary from the Alberta seasonal influenza outbreaks data. We simulate epidemics from the region restricted model (\ref{egdilmsim}).  Figure \ref{figh1} shows the progress of a typical epidemic over time under strong spatial dependency for a subset of the realizations from $t = 1$ up to $ t = 9$. The spatiotemporal structure of this epidemic shows that the outbreak of the epidemic is primarily driven by contacts within each LGA, followed by their neighboring LGAs. 

 Scenario $2$ (S$2$): We assume that $9$ individual DAs are infectious at time $t = 1$ (same as S$1$). We simulate epidemics from the global model (\ref{gdilm2sim}).  A typical epidemic over time under strong spatial dependency from $t = 1$ up to $ t = 9$ is shown in Figure $1$  of the Appendix in Section \ref{es}.

 Scenario $3$ (S$3$): An individual DA is randomly chosen to be infectious at the begining of the simulation $(t = 1)$. We  simulate epidemics from the global model (\ref{gdilm2sim}). A typical epidemic over time under strong spatial dependency from $t = 1$ up to $ t = 9$ is shown in Figure $2$ of the Appendix in Section \ref{es}.

Similar patterns of disease spread over time for other settings of the spatial dependency (moderate and weak) are observed in all three scenarios. Scenario $1$ is considered here to assess the performance of parameter estimates and inferences of the GD-ILMs when there is no mismatch between the data-generating mechanism and fitted model. However, when analyzing real epidemic outbreaks data, the transmission of the disease between individuals is not only limited to each LGA and their neighboring LGAs, it may happen across the entire study area. We therefore consider Scenarios $2$ and $3$ in which data are generated using the more realistic global generating model which allows for disease transmission between DAs across the whole study area. Scenarios $2$ and $3$ differ by varying infected DAs at the beginning of the simulation, allowing us to test if our results are robust to initial conditions.

\subsection{Model Fitting}\label{fitting}

The GD-ILMs defined in (\ref{egdilmsim}) was fitted to all simulated epidemics within a Bayesian framework via an MCMC algorithm, as described in Section \ref{mcmc}. The prior distributions for all model parameters were assumed to be independent,  and chosen to be weakly informative. More specifically, positive half-normal priors, each with mode $0$ and variance $100$, were used for $\alpha$, $\alpha_1$ and $\delta$ parameters.  Vague prior distributions for the spatial dependence parameter $\lambda$ was found to produce poorly mixing MCMC, thus a weakly informative hyperprior for $\lambda$ was used: $\lambda \sim ~ Beta(4, 2)$ (strong $\lambda$), $\lambda \sim ~ Beta(2, 2)$ (moderate $\lambda$), and $\lambda \sim ~ Beta(2, 4)$ (weak $\lambda$) \citep{macnab2014identification}. A  gamma prior was placed on the precision parameter $\tau = 1/\sigma^2$, $\tau \sim ~ Gamma(0.05, 0.05)$ and the LCAR prior distribution was used for the spatial random effects $\phi_k$.

A RWMH algorithm was used to update the $\alpha, \alpha_1, \delta, \phi_k$ and $\lambda$ parameters, with proposed values being drawn from normal distributions with proposal variances tuned to maintain an acceptance rate between $20\%$ and $50\%$. The variance parameter $\sigma^2$ was updated  using a Gibbs sampler from the full conditional inverse-gamma distribution. For each simulated epidemic, we ran a total of $300,000$ MCMC iterations with the first $50,000$ iterations discarded as burn-in, and we retained every $10$th sample in the remaining samples for inference. The convergence of MCMC was checked by visually inspecting their trace plots. 



\subsection{Results}\label{result}

\begin{figure}[!h]
\includegraphics[height = 2.7in, width = 6.5in]{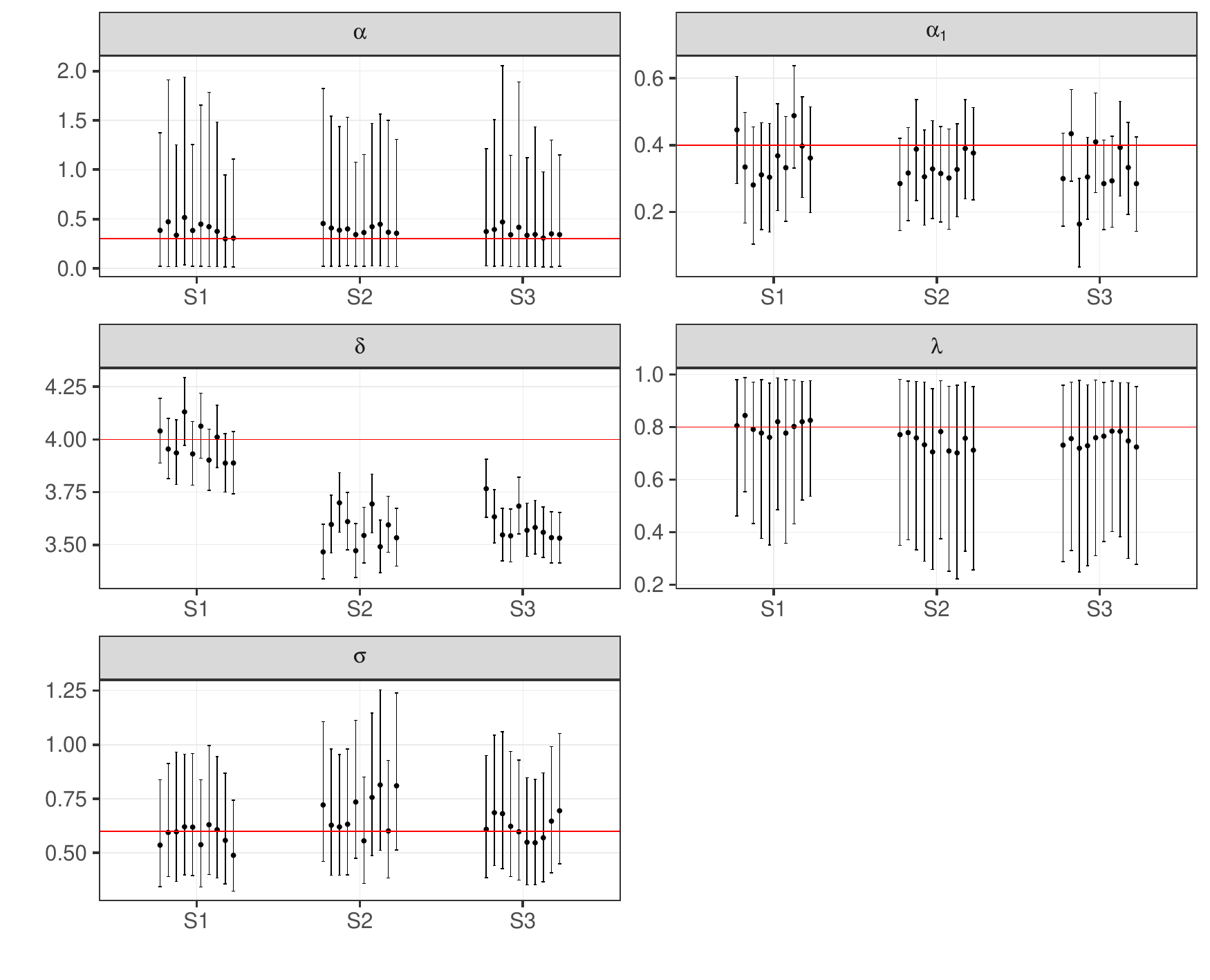}
\caption{Results of the fixed effects parameter for all Simulation Scenarios (S1-S3), under scenario of strong spatial dependency. The posterior mean estimates and $95\%$ CIs (vertical lines) for $\alpha, \alpha_1, \delta, \sigma$ parameters for each simulated epidemic where red lines indicate the true values.}
\label{figh4}
\end{figure}

\begin{figure}[!h]
\centering
\includegraphics[height = 2.7in, width = 6.5in]{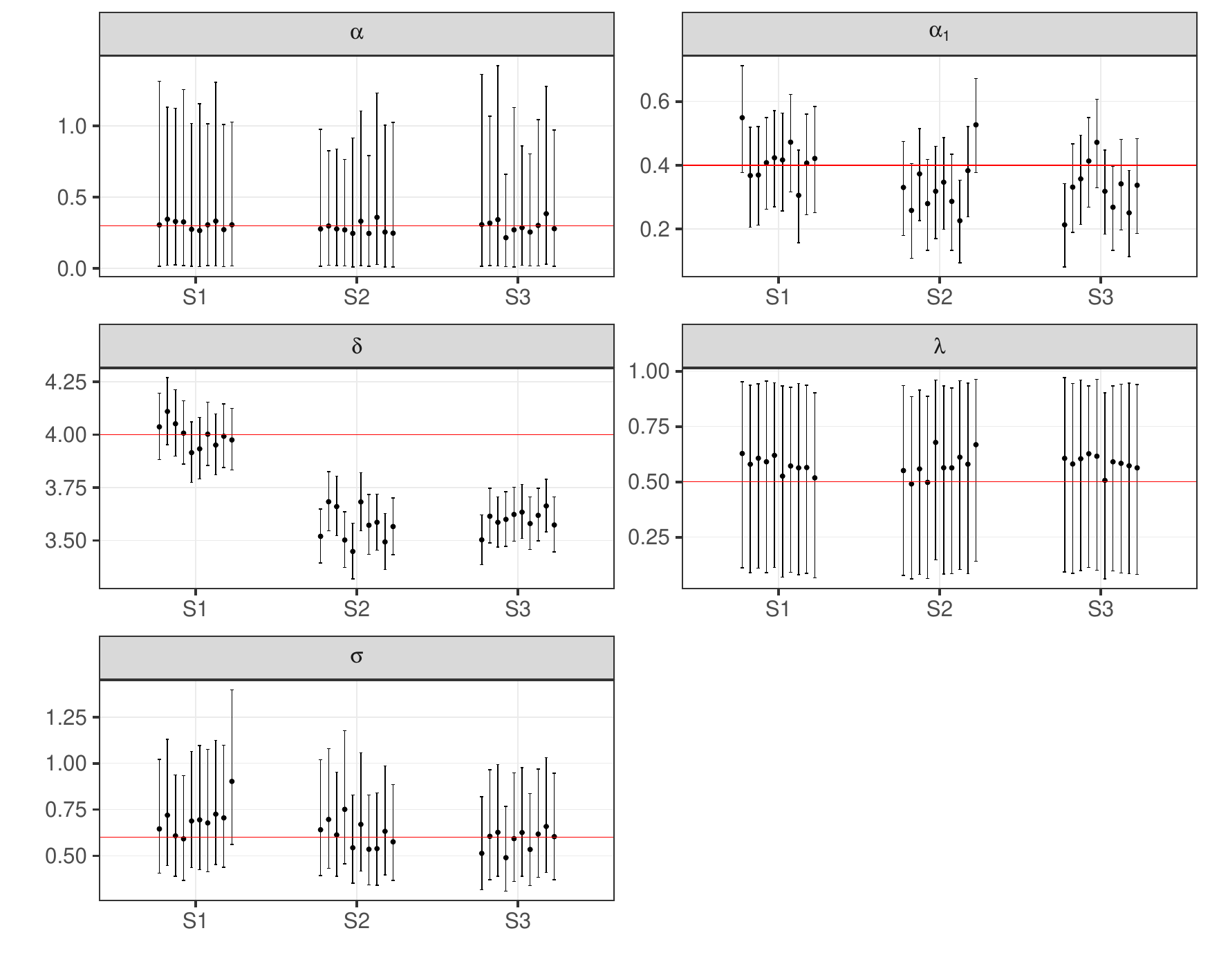}
\caption{Results of the fixed effects parameter for all Simulation Scenarios (S1-S3), under scenario of moderate spatial dependency. The posterior mean estimates and $95\%$ CIs (vertical lines) for $\alpha, \alpha_1, \delta, \sigma$ parameters for each simulated epidemic where red lines indicate the true values.}
\label{figm}
\end{figure}

\begin{figure} [!h]
\centering
\includegraphics[height = 2.7in, width = 6.5in]{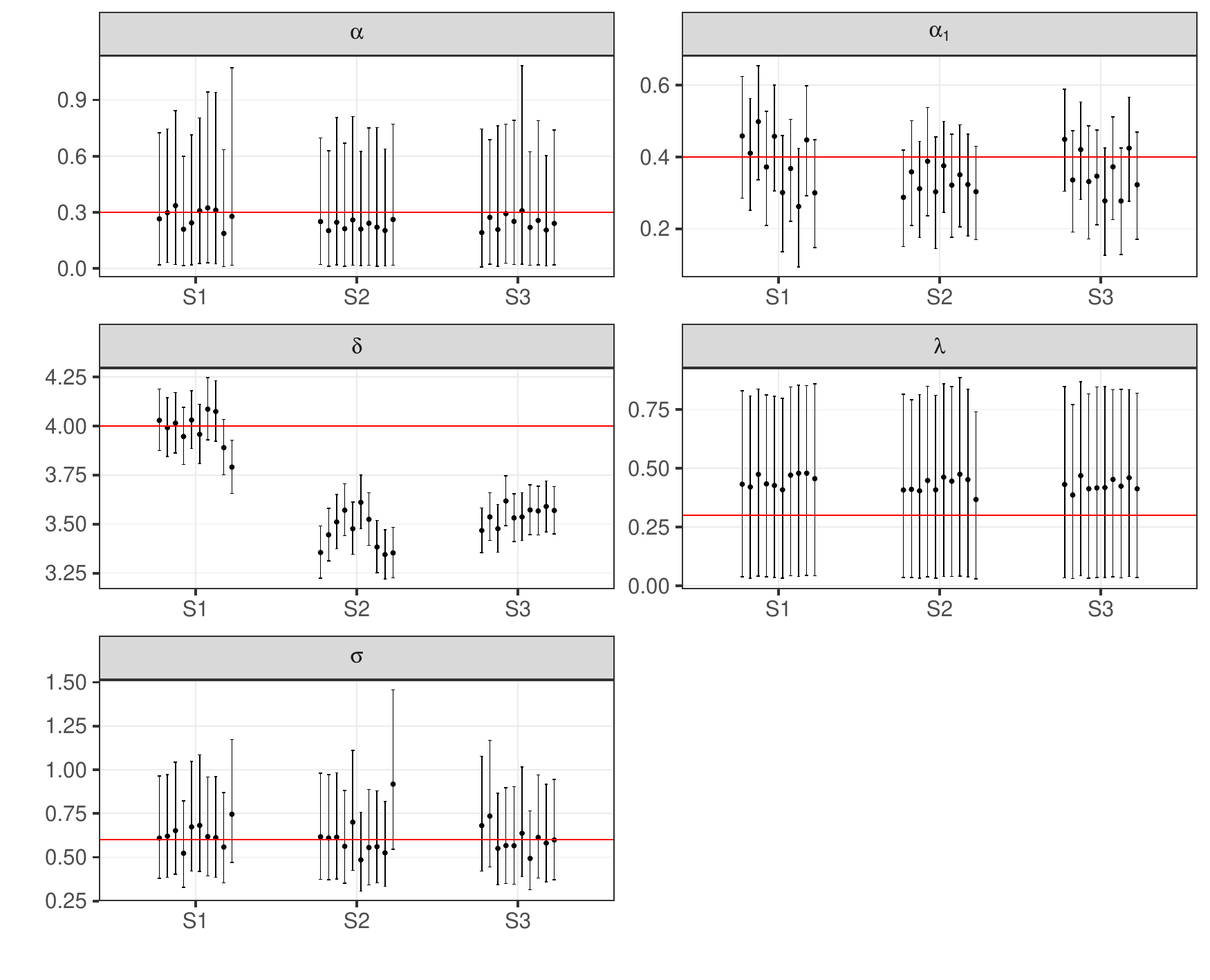}
\caption{Results of the fixed effects parameter for all Simulation Scenarios (S1-S3), under scenario of weak spatial dependency. The posterior mean estimates and $95\%$ CIs (vertical lines) for  $\alpha, \alpha_1, \delta, \sigma$ parameters for each simulated epidemic where red lines indicate the true values.}
\label{figw}
\end{figure}

Figure \ref{figh4} shows the posterior means and $95\%$ quantile-based  credible intervals (CIs) for each model parameter for each simulated epidemic in the case of strong spatial dependency, under Scenarios $1 - 3$. Under S$1$, all $95\%$ CIs overlap with the true parameter values used to generate the data, suggesting the GD-ILMs are able to recover appropriate parameter estimates when the fitted model has the same form as that used to generate the data.  Under S$2$ and S$3$, the posterior mean estimates are close to the true values for all model parameters with varying levels of uncertainty, except for  the spatial parameter, $\delta$, which is consistently underestimated.   This is understandable because  the susceptible individuals have much smaller sets of contactable infectious individuals under the fitted region restricted model than the generating global model. Thus, the infectious pressure on susceptible individuals will be smaller than under the true generating model with the same parameters. An underestimated $\delta$ compensates for this by increasing the probability of infections occurring by reducing the rate of decay of the power law spatial kernel. Figures \ref{figm} and \ref{figw} show the posterior means and $95\%$ CIs for each model parameter for the cases of moderate and weak spatial dependency, respectively. Once again, the true parameter values of all parameters fall within their respective $95\%$ CIs under all scenarios, except for $\delta$  which is underestimated under  S$2$ and S$3$. 

\begin{figure}[!h]
\centering
\includegraphics[height = 2.7in, width = 6.5in]{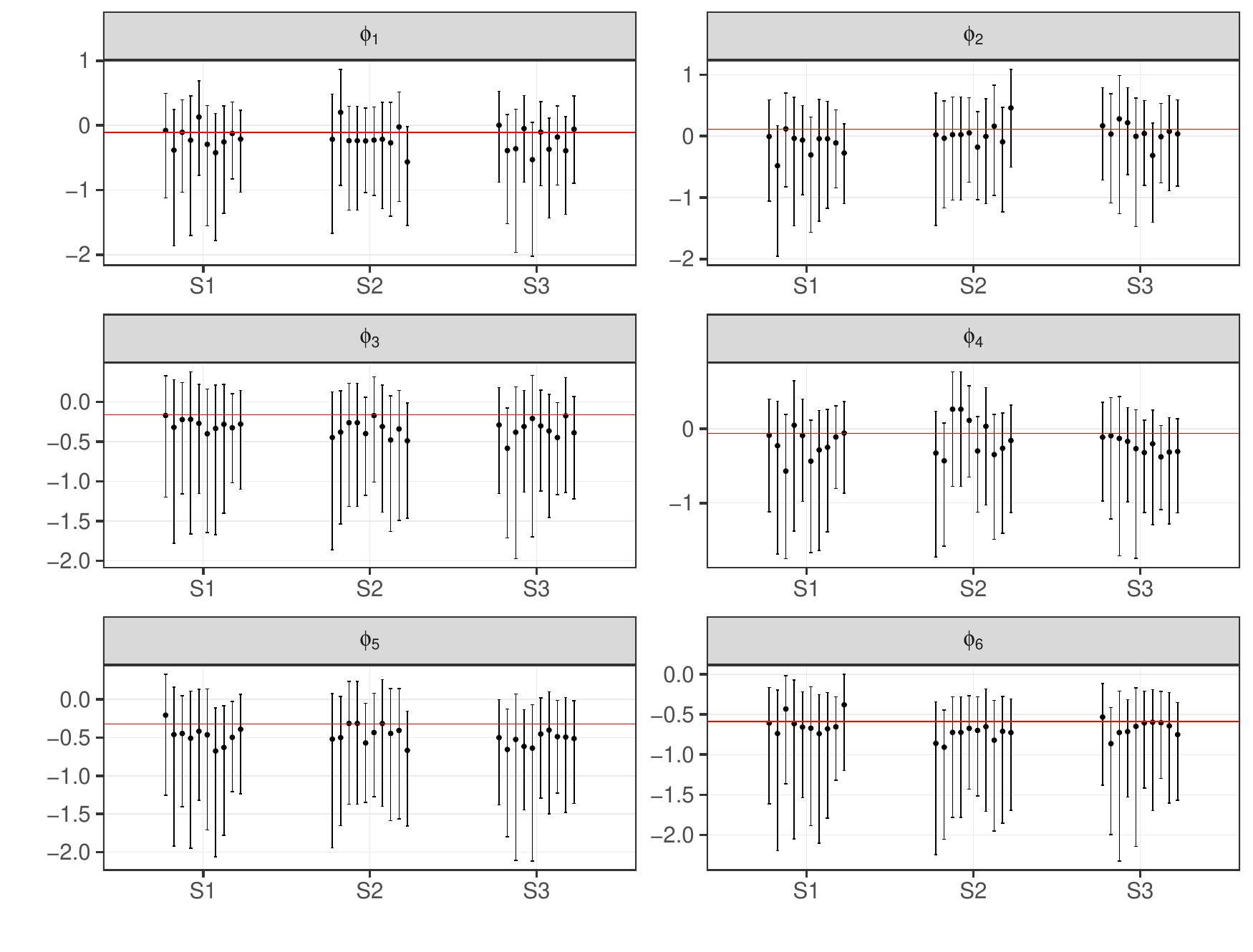}
\caption{Results of a subset of spatial random effects $(\phi_1 - \phi_6)$ across all Simulation Scenarios (S1-S3) in case of strong spatial dependency. The posterior mean estimates and $95\%$ CIs (vertical lines) for each simulated epidemic where red lines indicate the true values.}
\label{ran}
\end{figure}

Figure \ref{ran} shows the posterior means and $95\%$ CIs for a subset of the spatial random effects in the case of strong spatial dependency for a typical simulated epidemic under S$1$-S$3$. The posterior mean estimates of the spatial random effects are close to their true values under all scenarios, albeit with varying levels of posterior uncertainty.  This was also the case for other settings of spatial dependency (results not shown). Thus, the spatial heterogeneity was successfully captured by the spatial random effects.

\begin{figure}[!h]
\centering
\begin{subfigure}[b]{0.45\textwidth}
\includegraphics[width=\textwidth]{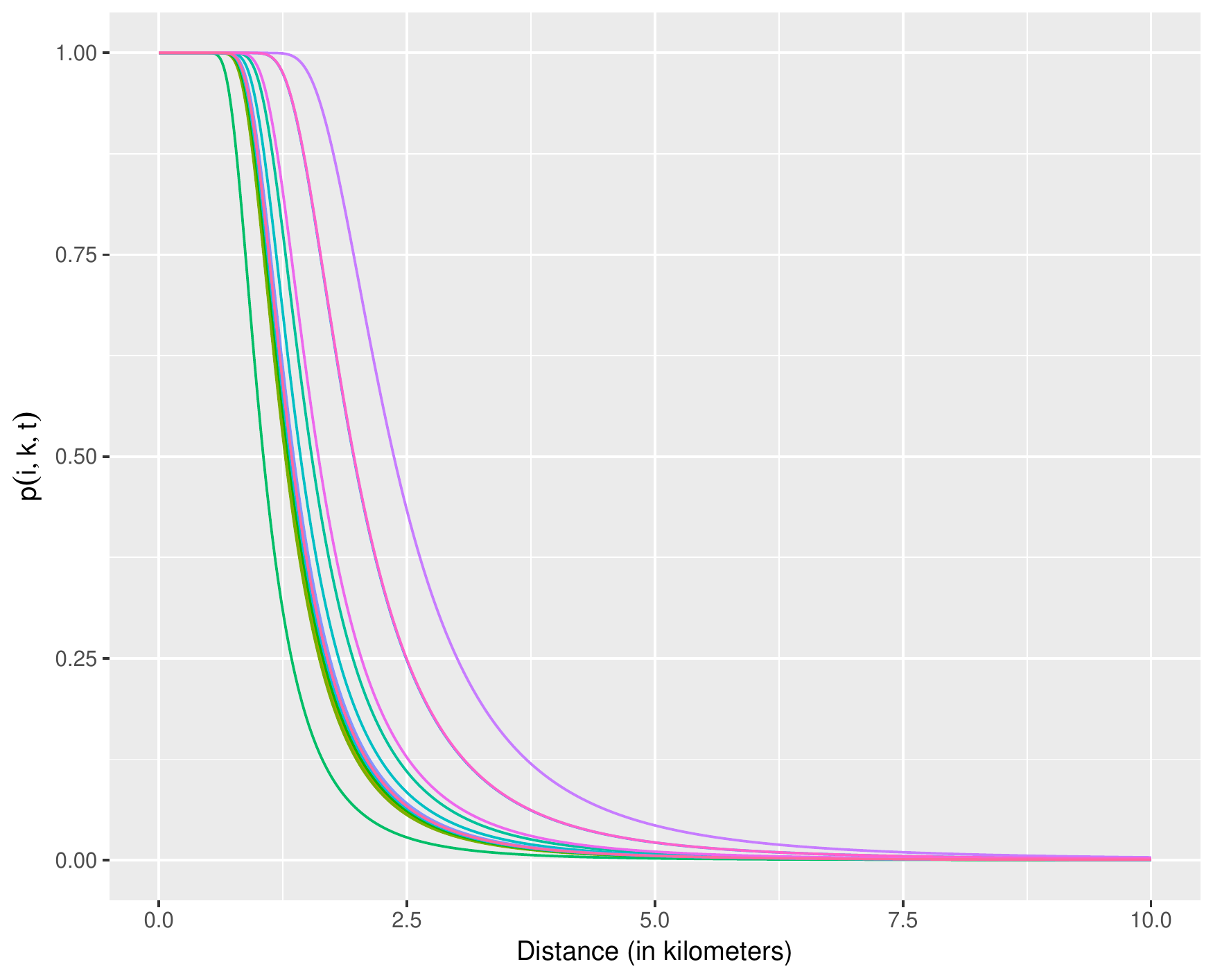}
\caption{$P(i, k, t)$ vs distance}
\label{pinf}
\end{subfigure}
~
\begin{subfigure}[b]{0.45\textwidth}
\includegraphics[width=\textwidth]{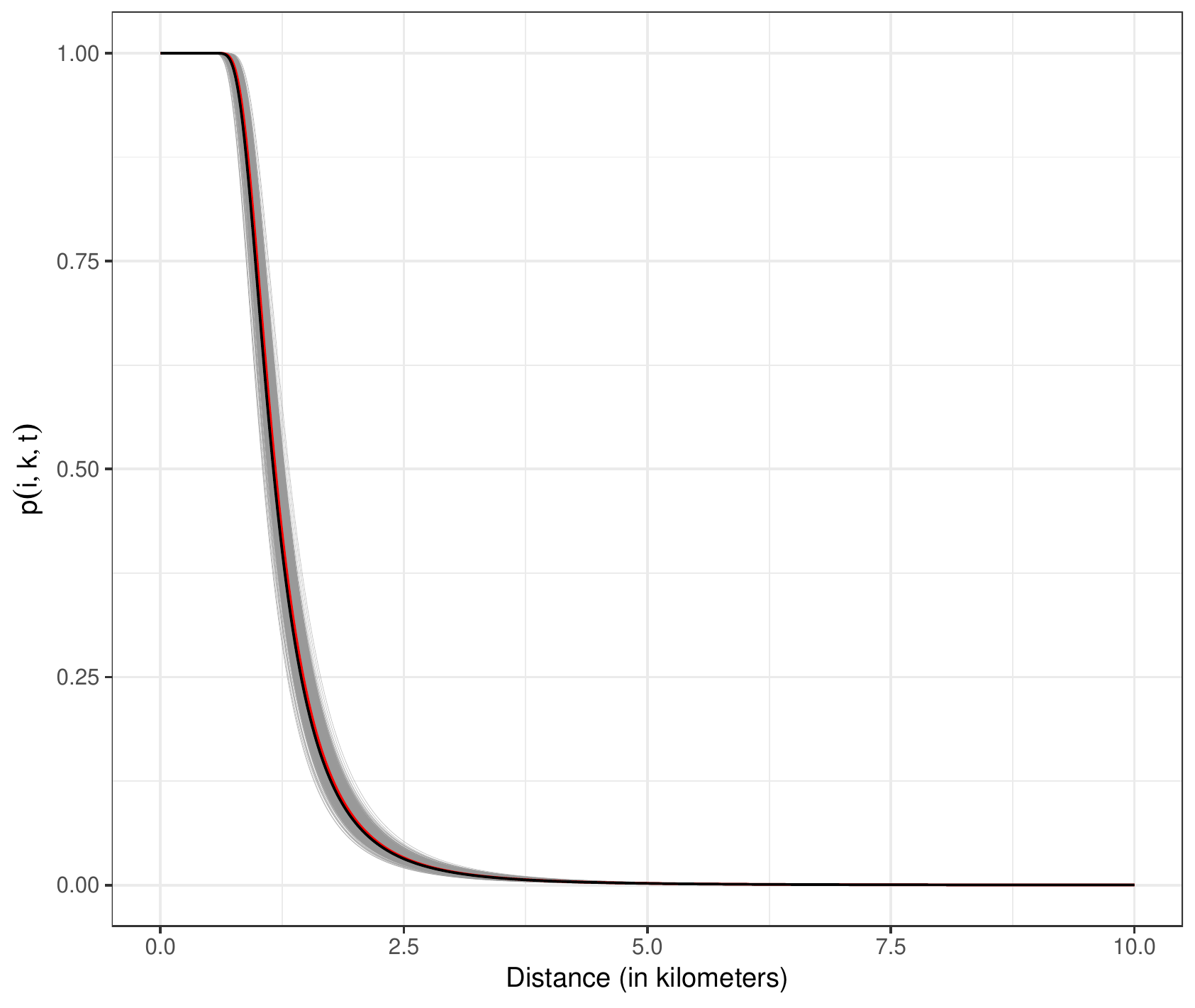}
\caption{$\kappa(i, j)$ vs distance}
\label{pker}
\end{subfigure}
\caption{ The probability of susceptible DA $i$ being infected from infectious DA $j$ within a time unit using median population size of DAs for each LGA against distance (in kilometers), (a) under the posterior mean where each line representing a different LGA; and (b) the gray lines represent a random sample of  $1000$ posterior samples, and the red (black) line represents posterior mean (true values) for the West Bow LGA.}
\label{sker}
\end{figure}

The spatial signal identified (here, for a typical epidemic in Scenario $2$ under strong spatial dependency setting) is shown in Figure \ref{sker}. In particular, Figure \ref{pinf} shows the posterior mean of the probability of a susceptible DA $i$ being infected from an infectious DA $j$ over distance ($d_{ij}$), for each of the $16$ LGAs.   Figure \ref{pker} shows the posterior predictive distribution of the probability of infection against distance (gray lines), based on a random sample of $1000$ posterior samples, with the posterior mean (red line), and true values (black line) for the West Bow LGA. We can see that the posterior mean of the random sample follows the true course of the epidemic very closely, and that the variation of curve under the posterior is low. Similar results are seen for different LGAs. The variation between the curves in Figure \ref{pinf} shows the effect of the spatial random effects on one-to-one DA infection over distance. Overall, these results imply that the spatial spread of disease in our simulation setting, is very low over distances above $5.0$ km.

\section{Application to Alberta Seasonal Influenza  Outbreak Data} \label{rdata}

We now apply  $\mathcal{SI}$ and $\mathcal{SIR}$ compartmental frameworks in the region-restricted GD-ILMs to data on the seasonal influenza outbreak that occurred during the period of October $25$ - November $14, 2009$ in Calgary, Canada. However, in this case it turns out that the epidemic has already peaked. This would be in line with a scenario in which we aim to model the influenza epidemic in order to use the fitted model for forecasting, and/or quantifying areas with higher infectivity rates to the development of disease control.  

\subsection{Data Description}

\begin{figure}[!h]
\centering
\includegraphics[height = 5.0in, width = 6.5in]{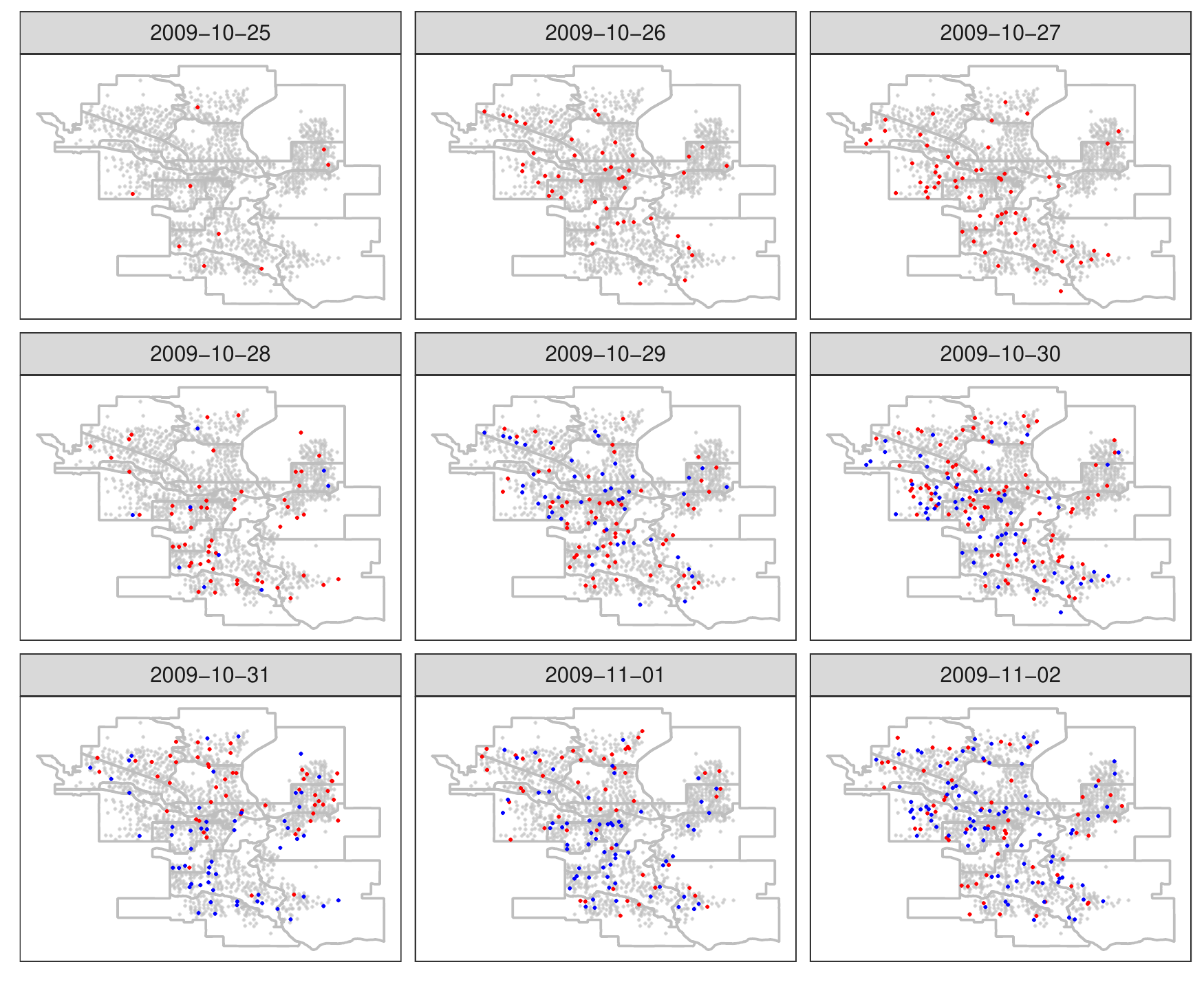}
\caption{A subset of the influenza epidemics during the period of October $25$ - November $2, 2009$ across the $16$ LGAs for the city of Calgary, Canada. Newly infected, susceptible, and removal individual DAs are denoted by red, grey and blue dots, respectively.}
\label{exreal}
\end{figure}

Data on  daily physician visits  due to influenza with pneumonia (ICD-9: 487.0), influenza with other respiratory manifestations (ICD-9: 487.1), and influenza with other manifestations (ICD-9: 487.8) were obtained from the Alberta Health, Analytics and Performance Reporting Branch of the Government of Alberta. Data on the residential postal code of each patient identified under one of these categories was also obtained. The \citet{AHS} postal code translator file was used to link each patient six-character postal code to their DA and LGA. An individual DA is considered to become infectious on the first day that a patient within that DA is diagnosed as having influenza according to one of three physician diagnosed categories above.   We defined an individual DA as susceptible on a given day if no patient had visited physician up to that point in time. Under the $\mathcal{SIR}$ framework DAs are only assumed to become infectious once in the three week period considered.  Figure \ref{exreal} presents a map of the infected DAs under  the $\mathcal{SIR}$ framework (assuming an infectious period of three days) during the period of October $25$ - November $2, 2009$.  A total of $9$ individual DA in $6$ LGAs were found infectious on the  October $25, 2009$, as discussed in Section \ref{episimu}. Note that, $912 (58.1\%)$ of $1570$ DAs were infected during the three week period of seasonal influenza epidemics being modelled.

\subsection{Data Analysis}

Two compartmental frameworks were fitted for the seasonal influenza outbreak data in Calgary, Canada. One was an  $\mathcal{SIR}$ framework in which the infectious period was assumed to be $\gamma = 3$ days for all infected DAs. The second was an $\mathcal{SI}$ framework  which assumed infectiousness lasted for the whole study period after infection. The $\gamma = 3$ assumption under the  $\mathcal{SIR}$ framework is akin to what is known about human-level dynamics. However, we are fitting our models at the DA level, and so  $\gamma = 3$ assumption would seem as naive (akin to assuming that only one person per DA is infected). The $\mathcal{SI}$ structure would seem to be more reasonable for DA-level modeling. However, we include both analyses here in order to investigate the robustness of results to the infectious period assumption.

The  $\mathcal{SI}$ and $\mathcal{SIR}$ frameworks in the region-restricted GD-ILM defined in (\ref{egdilmsim})  were fitted to the $2009$ Calgary seasonal influenza outbreak data using the MCMC procedure described in Section \ref{mcmc}.  The constant infectivity parameter ($\alpha$) was set to zero in the fitted model as it was found to lead to better mixing of the spatial random effects parameters and lead to faster MCMC convergence. (Note, however, it is perfectly possible to fit the model with this parameter included).

The prior distributions of the model parameters in (\ref{egdilmsim}) were assumed same as simulation study that described in Section \ref{fitting}, with an exception on the spatial dependency parameter which was assigned a uniform prior, $\lambda \sim U(0, 1)$. A total of $150,000$ MCMC iterations were run. The first $50,000$ iterations were discarded as burn-in, and every $10^{th}$ sample was retained in the remaining samples for inference. The convergence of MCMC was checked by visually inspecting the trace plots. 
\subsection{Results}

Posterior means and quantile-based CIs for all model parameters under $\mathcal{SI}$ and $\mathcal{SIR}$ frameworks are presented in Table \ref{post}.  Under both of these framework, it was estimated that population size has a positive effect on the influenza transmission, so a highly populated DA is more likely to promote the spread of virus to other DAs than the one with a low population. However, this effect was seen to be notably larger under the $\mathcal{SI}$ framework ($0.933$) than the $\mathcal{SIR}$ framework ($0.714$), with no overlap between the $95\%$ CIs.   The posterior means were also determined for the spatial random effects and they ranged from $- 5.91$  to $- 4.44$, and $- 4.61$ to $- 3.24$, across the $16$ LGAs for the $\mathcal{SI}$ and $\mathcal{SIR}$ frameworks, respectively. The posterior results for the other model parameters were similar under the $\mathcal{SI}$ and $\mathcal{SIR}$ frameworks. Here,  we focus on results under the  $\mathcal{SIR}$ framework. 

\begin{table}[!h]
\caption{Posterior means and $95\%$ credible intervals for the parameters in the GD-ILMs under $\mathcal{SI}$  and $\mathcal{SIR}$ frameworks}

\begin{tabular}{l c c c c c}
\hline 
&  &  $\mathcal{SI}$  & &  & $\mathcal{SIR}$ \\ \cline{2 - 3} \cline{5 - 6}
Parameter & Mean & $95\%$ credible interval & & Mean & $95\%$ credible interval \\
\hline
$\alpha_1$ & 0.933 & (0.837, 1.023) &  & 0.714 & (0.619, 0.807)    \\
$\delta$ & 0.149 & (0.011, 0.334) &    &  0.134 & (0.012, 0.288) \\
$\lambda$ & 0.982 & (0.976, 0.999) &  & 0.986 & (0.974, 0.996) \\
$\sigma$ & 1.064 & (0.694, 1.632) &  & 0.985 & (0.636, 1.519)\\
\hline

\end{tabular}
\label{post}
\end{table}

\begin{figure}[!h]
\centering
\begin{subfigure}[b]{0.45\textwidth}
\includegraphics[width=\textwidth]{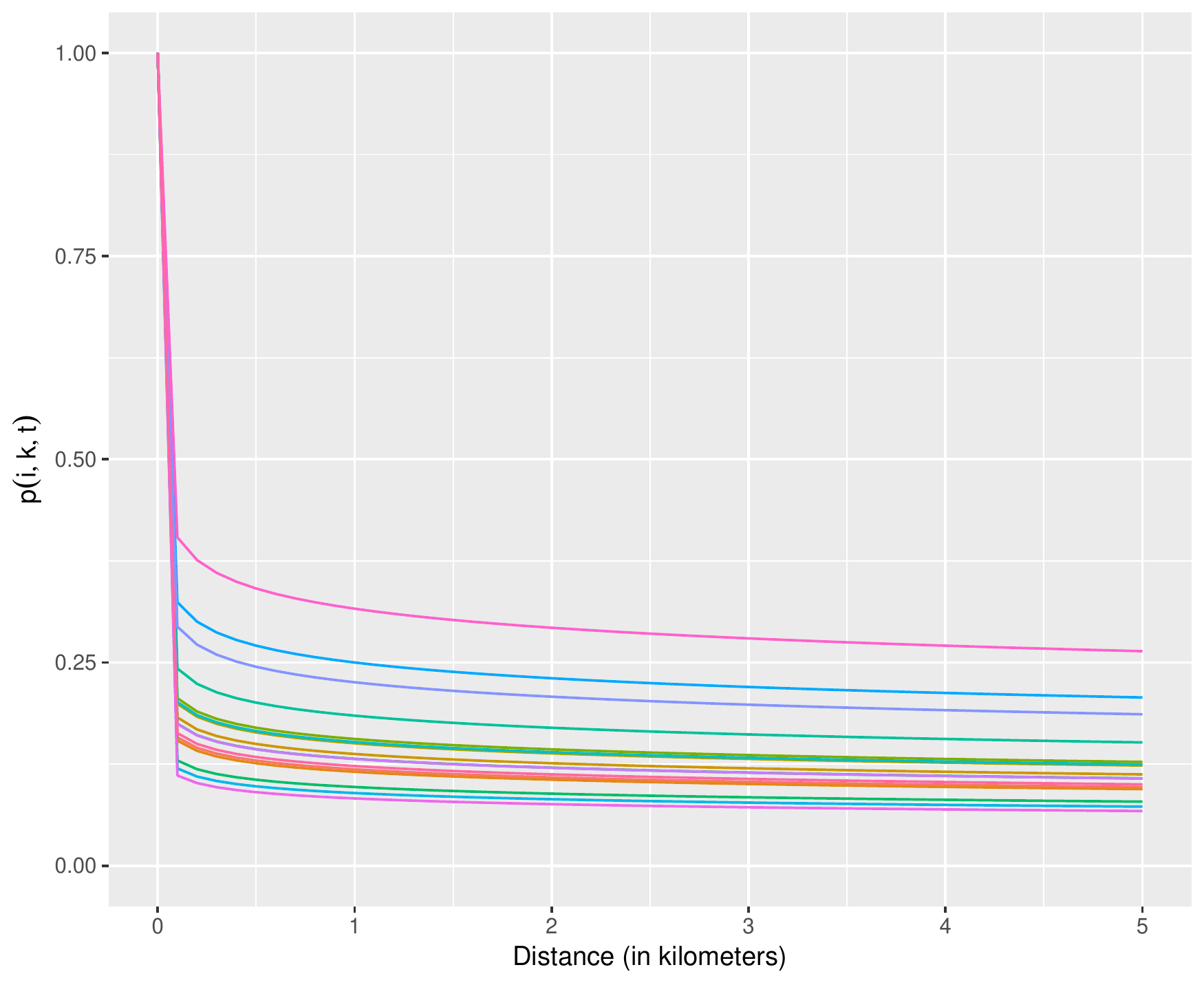}
\caption{$P(i, k, t)$ vs distance}
\label{rker1}
\end{subfigure}
~
\begin{subfigure}[b]{0.45\textwidth}
\includegraphics[width=\textwidth]{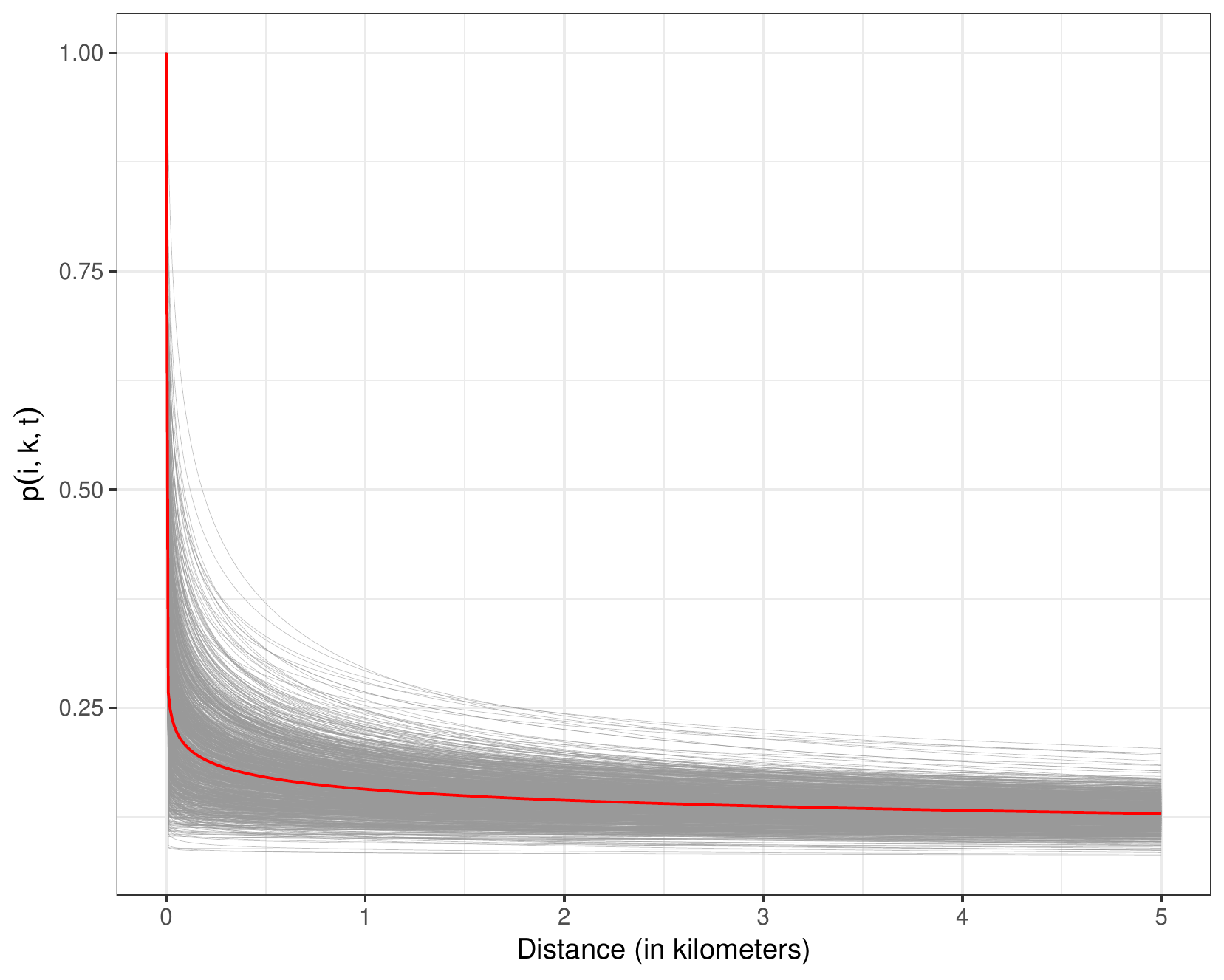}
\caption{$P(i, k, t)$ vs distance}
\label{rker2}
\end{subfigure}
\caption{The probability of susceptible DA $i$ being infected from infectious DA $j$ across the $16$ LGAs using   median population size of DAs for each LGA against distance (in kilometers), (a) under the posterior mean where each line representing a different LGA; and (b) the gray lines represent a random sample of $1000$ posterior samples, and the red line represents posterior mean for the West Bow LGA.}
\label{ker}
\end{figure}

Figure \ref{ker} illustrates the  probability of infection with spatial distance kernel. Figure \ref{rker1} shows that the probability of susceptible DA $i$ being infected from a single infectious DA $j$ over distance based on median population size of DA within each LGA. Figure \ref{rker2} shows the posterior predictive distribution of the probability of infection over distance (gray lines), based on a random sample of $1000$ posterior samples, with the posterior mean (red line) for the West Bow LGA. We can see that the posterior predictive infection probability of the random sample is close to their posterior mean with varying levels of uncertainty and similar results are seen for other LGAs. Although the posterior predictive variance is quite high, it appears that spatial distance was not an important factor in the transmission of influenza, with posterior mean estimates that lead to fairly flat curves over meaningful distances.  This would imply that, at least using data from the beginning of an outbreak, spatial distance between DA centroids is not a good predictor of the first infectious in DAs.  Overall, then it would seen that population size and LGA-level random effects play more of a role in the local spread of seasonal influenza.

\begin{figure}[!h]
\centering
\includegraphics[height = 5.0in, width = 6.5in]{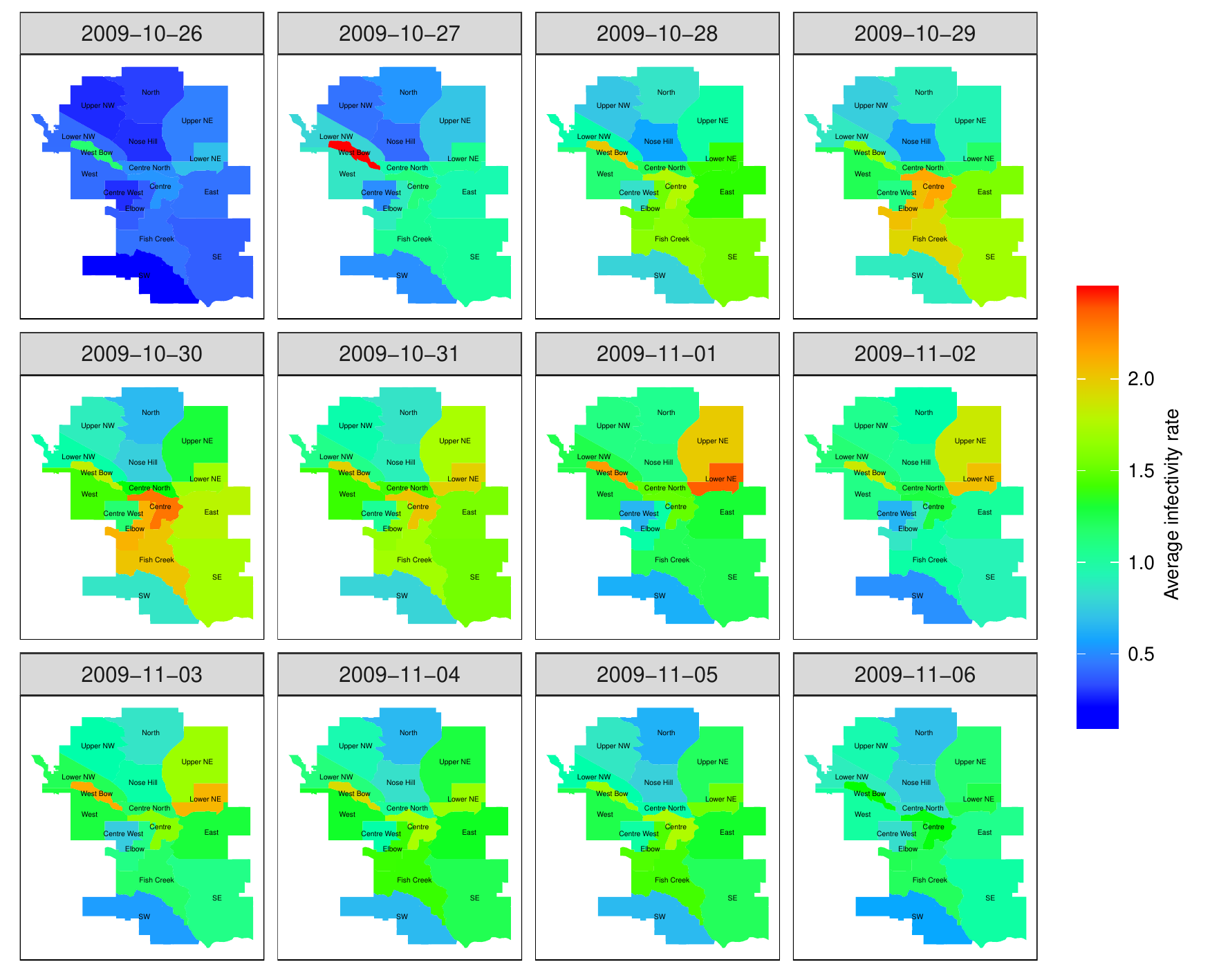}
\caption{ The spatiotemporal distribution of posterior mean infectivity rates of influenza outbreaks during the period of October $26$ - November $6, 2009$ across the $16$ LGAs  for the city of Calgary, Canada.}
\label{rinfrate}
\end{figure}

Finally, in order to make the results of inference more directly relevant to the development of disease surveillance and control, we construct posterior distribution of infectivity rates that quantify the LGA-level infection risk over time. The posterior mean infectivity rate for each susceptible individual DA was determined based on the posterior mean of the fixed and spatial random effects parameters and then summarized it by averaging over the susceptible DAs as function of time. The average of the posterior mean infectivity rates can be used to produce risk maps for showing the spatial distribution of LGA-level infection risk at a given point in time. These maps may be used to inform targeted surveillance by ranking LGAs in order of the most likely to be infected. Figure \ref{rinfrate} showed a subset of the risk maps for influenza outbreaks during the period of October $26$ - November $6, 2009$ across the $16$ LGAs  for the city of Calgary, Canada.

The results in Figure \ref{rinfrate} show that there were striking  differences in the spatial variation of the posterior mean daily infectivity rate during the early epidemic period (October $26$, $2009$) and peak epidemic period (October $30$, $2009$). In addition, the spatiotemporal structure of the posterior mean infectivity rates imply that the influenza outbreaks may be  being  driven to a large degree by contacts within LGAs as well as to their neighboring LGAs. In the peak epidemic period, the LGAs:  Centre ($2.24$),  Elbow ($2.03$), and  Fish Creek ($1.96$), experienced highest posterior mean infectivity rate while  West Bow ($1.76$),  East ($1.73$), SE ($1.66$), and Lower NE ($1.63$), experienced the next highest level of infection risk. It was worth noting that the average of the posterior mean infectivity rates for densely populated LGAs were largest.

\section{Discussion}\label{discuss}

In this paper, we have extended the framework of \citet{deardon2010inference} to enable inferences to be made about infectious disease dynamics with respect to time and space while accounting for unobserved spatially structured latent covariates. The resulting GD-ILMs were tested using simulated epidemic data, and by application to data from a seasonal outbreak of influenza in Calgary, Canada. We would expect that our approach can be applied to a wide range of infectious disease epidemics for humans and animals(e.g., foot and mouth disease, zika virus, ebola virus and avian influenza), and become a highly useful tool in infectious disease epidemiology.

Analysis under a number of simulated epidemics scenarios showed that inferred parameter estimates were reliable with the true parameters falling within their respective $95\%$ credible intervals. This was even the case when fitting the region-restricted model to data produced by a non-restricted model, except for the spatial transmission kernel parameter $(\delta)$, which tended to be underestimated in such situations. We have shown new GD-ILMs can be used to conduct risk assessments via posterior infectivity rate maps that account for spatial structure.  Such risk maps can be used to inform disease control efforts, for instance, by targeting vaccination efforts in high risk LGAs, and/or providing extra resources for hospital emergency departments in those (or nearby) LGAs.

We have illustrated through simulation that the GD-ILMs can capture both spatial transmission dynamics within regions through the infection kernel, and regional spatial heterogeneity. However, in the context of seasonal influenza outbreak data, we observed that spatial distance did not appear to be an important factor in the transmission of seasonal influenza between DAs in Calgary in 2009. On the other hand, the population size and regional-level random effects did appear to be important factors in the transmission of seasonal influenza.  This may be expected since humans tend to have high mobility, and may well be travelling large distances throughout the city on a daily basis. Note, we also analyzed the seasonal influenza data using the Cauchy distance kernel of \citet{jewell2009bayesian} instead of using power law kernel and still did not find any spatial signal in the transmission of seasonal influenza.

There are a number of potential avenues for future work. A useful feature of the Bayesian MCMC framework is that it is well suited to handling the challenges that often arise in epidemic modeling due to the partial nature of observations, and allows unobserved quantities (e.g., time of infection) and removal (e.g., recovery or death) to be accommodated in analyses using data-augmentation treating the infection and removal times as latent variables. However, fitting GD-ILMs  without allowing for event time uncertainty, as we have done here, is already computationally intensive. That said,  part of our ongoing work is to incorporate such uncertainty into the analysis, and explore alternative computational approaches. For example,  a focus of future work could be to use approaches such as approximate Bayesian Computation (ABC) that attempt to avoid explicit calculation of the likelihood function \citep{marjoram2003markov, beaumont2009adaptive}. This has been done in an infectious disease modeling environment for simple homogeneous models \citep{mckinley2014simulation, mckinley2009inference}, as well as in an ILM framework \citep{almutiry2018incorporating}.  Alternatively, the data-sampling likelihood approaches of \citet{malik2016parameterizing}, or Gaussian process emulation methods of \citet{pokharel2016gaussian}, both of which have been implemented for spatial ILMs, might also be interesting paths to go down.

Finally, the modeling framework can be easily  extended to allow for more complicated disease life histories than that of the $\mathcal{SI}$ or $\mathcal{SIR}$ framework.  For instance, an $\mathcal{SEIR}$ framework can be considered in which individuals enter a latent period before becoming infectious after exposure to the disease. More pertinently for a disease such as influenza being modelled at the DA scale, it would be desirable to extend the compartmental framework to allow for reinfection (e.g., an $\mathcal{SIRS}$ or $\mathcal{SEIRS}$ framework).

\section*{Acknowledgments}
We thank Larry Svenson and others at the Alberta Health for providing the Alberta seasonal influenza outbreak data (2009). We also thank Vineet Saini at the Alberta Health Services for providing the LGA boundaries shapefiles. Funding for this study was provided by the Canadian Statistical Sciences Institute (CANSSI) under collaborative research team project and Natural Sciences and Engineering Research Council (NSERC) discovery grant. \\
\section*{Conflict of Interest} The authors declare that they have no conflict of interest.

\bibliography{reference}

\pagebreak
\begin{appendices}

\section{Conditional autoregressive models}\label{lcar}
A single set of spatial random effects $\Phi = (\phi_1, \phi_2, ....,\phi_K)$, which are represented by the multivariate normal distribution:
\begin{equation}\label{lcar}
\Phi \sim MVN \left(\mathbf{0}, \sigma^2 \left[\lambda \mathbf{R} + (1- \lambda) \mathbf{I} \right]^{-1} \right),
\end{equation}
\noindent where $\sigma^2$ is the parameter controlling the variance of random effects, $\lambda$ is a spatial dependence parameter lying in the interval $[0,1]$, and the $k\ell$th element of $\mathbf{R}$ is defined as
\begin{equation}\label{dvar}
R_{k\ell} = 
\left\{
\begin{aligned}
& m_{k} ~~~~ \ell = k ,\\
& -1 ~~~ \ell \sim k ,\\
& 0 ~~~\text{otherwise},
\end{aligned}
\right.
\end{equation}

\noindent where $m_k$ is the number of neighbors of region $k$, $\ell \sim k$ indicates that regions $\ell$ and $k$ are neighbors.

\noindent The precision matrix $\mathbf{L} = \lambda \mathbf{R} + (1- \lambda) \mathbf{I}$, where $\mathbf{I}$ denotes an identity matrix of order $K $(number of areas), is positive-definite and symmetric and is a weighted average of spatially dependent (represented by $\mathbf{R}$) and independent (denoted by $\mathbf{I}$) correlation structures, where the weight is equal to $\lambda$. The strength of the spatial autocorrelation is controlled by $\lambda$, with $\lambda = 0$ yields the independence case $(\mathbf{L} = \mathbf{I})$ and intrinsic CAR $(\mathbf{L} =  \mathbf{R})$ for $\lambda = 1$. If $0 \le \lambda < 1$, the joint distribution (\ref{lcar}) is proper, while $\lambda =1$ corresponds to the improper intrinsic CAR model. The univariate full conditional distributions corresponding to (\ref{lcar}) are given by
\begin{equation}\label{lcareq}
\phi_k|\phi_{-k} \sim  N\left(\frac{\lambda}{m_k\lambda + 1 -\lambda}\displaystyle\sum_{\ell \sim k }^{} \phi_{\ell} , \frac{\sigma^2}{m_k\lambda + 1 - \lambda} \right),
\end{equation}

\noindent where $\phi_{-k} = (\phi_1, ....,\phi_{k -1}, \phi_{k + 1}, ...,\phi_K) $, that is, the random effect vector with the $k$th component deleted. The conditional mean can be seen as a weighted average of the random effects in neighboring areas (with weight $m_k \lambda $) and the overall average $0$ (with weight $1 - \lambda$). The conditional variance can similarly be written as a weighted average of the local variance from the intrinsic autoregression and the variance from the independence model.

\section{Epidemic Simulation}\label{es}

\begin{figure}[!h]
\centering
\includegraphics[height = 5.0in, width = 6.5in]{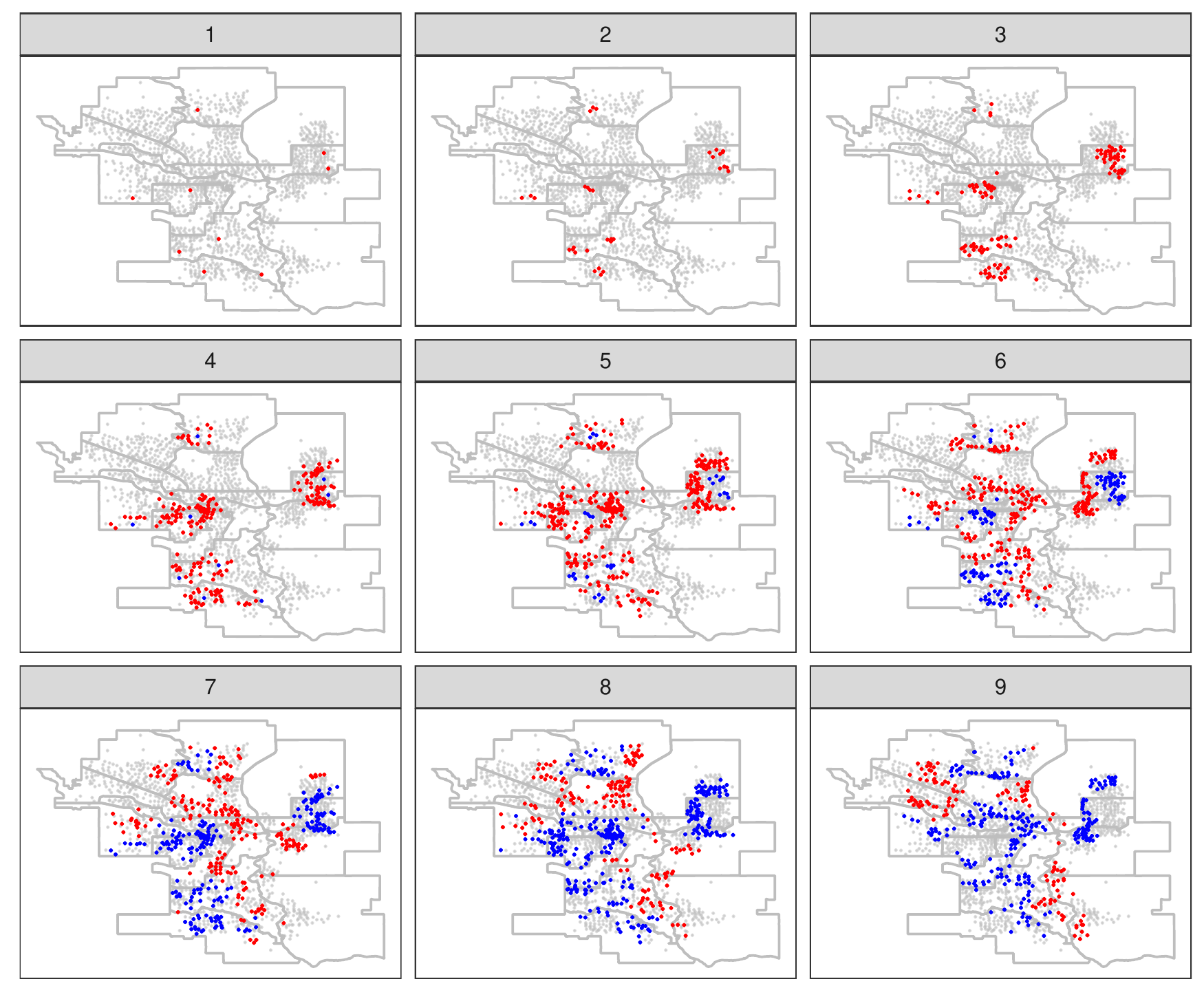}
\caption{A subset of a realization of the epidemic progress maps made from $t = 1$ up to $ t = 9$ are shown for the city of Calgary, Canada under the Scenario $2$ (S$2$). Newly infected, susceptible, and removal individual DAs are denoted by red, grey and blue dots, respectively.}
\label{fighs2}
\end{figure}

\begin{figure}[!h]
\centering
\includegraphics[height = 5.0in, width = 6.5in]{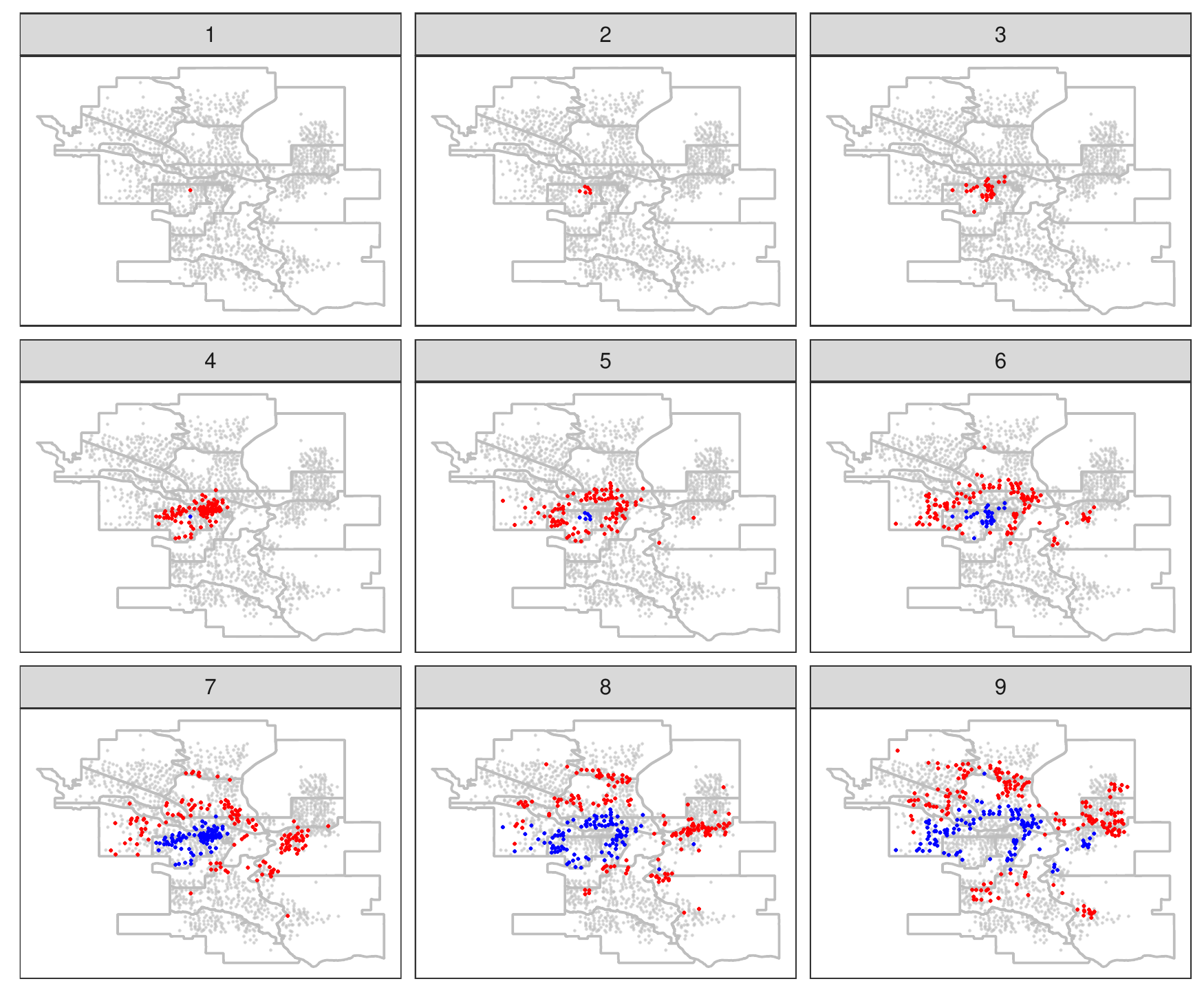}
\caption{A subset of a realization of the epidemic progress maps made from $t = 1$ up to $ t = 9$ are shown for the city of Calgary, Canada under the Scenario $3$ (S$3$). Newly infected, susceptible, and removal individual DAs are denoted by red, grey and blue dots, respectively.}
\label{fighs3}
\end{figure}

\end{appendices}

\end{document}